# Social Computing for Mobile Big Data in Wireless Networks[1]

**Xing Zhang[12], Zhenglei Yi[1], Zhi Yan[3], Geyong Min[4], Wenbo Wang[12], Sabita Maharjan[5], Yan Zhang[5]**

1 Beijing Advanced Innovation Center for Future Internet Technology, Beijing University of Technology, Beijing, China

2 Key Laboratory of Universal Wireless Communication, Ministry of Education, Beijing University of Posts and Telecommunications, Beijing, 100876, China, Email: zhangx@ieee.org

3 School of Electrical and Information Engineering, Hunan University, Changsha, China.

4 College of Engineering, Mathematics and Physical Sciences, University of Exeter, U.K.

5 Simula Research Laboratory, Fornebu 1364, Norway

**Abstract:** Mobile big data contains vast statistical features in various dimensions, including spatial, temporal, and the underlying social domain. Understanding and exploiting the features of mobile data from a social network perspective will be extremely beneficial to wireless networks, from planning, operation, and maintenance to optimization and marketing. In this paper, we categorize and analyze the big data collected from real wireless cellular networks. Then, we study the social characteristics of mobile big data and highlight several research directions for mobile big data in the social computing areas.

## 1. Introduction

Data services' exponential growth, and the constantly expanding wireless and mobile applications that use them, have ushered in an era of big data. Since 2014, the number of connected mobile devices has been more than the world's population. The surge of mobile traffic in recent years is mainly attributed to the rapid proliferation of mobile social applications including multimedia, running on mobile devices such as smartphones, mobile tablets, and other smart mobile devices. By 2020, more than three-fifths of all devices connected to the mobile networks will be "smart" devices. With a compound annual growth rate (CAGR) of 53% global mobile data traffic will increase nearly eightfold between 2015 and 2020 [1].

This mobile big data poses many new challenges to conventional data analytics because of its large dimensionality, heterogeneity, and complex features therein, e.g., Volume, Variety, Velocity, Value and Veracity [2]. Dealing with big data is a key challenge for many wireless networking applications such as O&M, planning and optimization, and marketing. In this paper, to help address this problem, we provide a

---

classification structure for mobile big data and highlight several research directions from the perspective of social computing using a significant volume of real data collected from mobile networks.

## 2. Big Data in Mobile Cellular Networks

The concept of 'Big Data' means not only a large volume of data but also other features that differentiate it from the concepts of `huge amount data'. The big data definition given in [2] includes the 5V properties: Volume, Variety, Velocity, Value and Veracity. It's a new generation of technologies and architectures designed to economically extract value from very large volume of a wide variety of data by enabling high velocity capture, discovery, and analysis. It contains massive volume of both structured and unstructured data that is difficult to process using traditional database and software techniques.

In addition to the five 'Vs' properties, due to the complexity of mobile cellular networks, big data in mobile cellular networks also exhibits several other unique characteristics, which lead to unprecedented challenges as well as opportunities . For instance, to understand behaviors and requirements of mobile users, which in turn allow the intelligent decision making for real-time decision making in various    applications. In this section, we will introduce the data categories in mobile cellular networks and their unique characteristics.

## 2.1 Data Categories in Mobile Cellular Networks

The vast amount of mobile data is collected and extracted from several key network interfaces, in both Radio Access Network (RAN) and Core Network (CN), as shown in Figure 1. These data can be roughly classified into four categories that include flow record data, network performance data, mobile terminal data, and additional data information, as shown in Table 1.

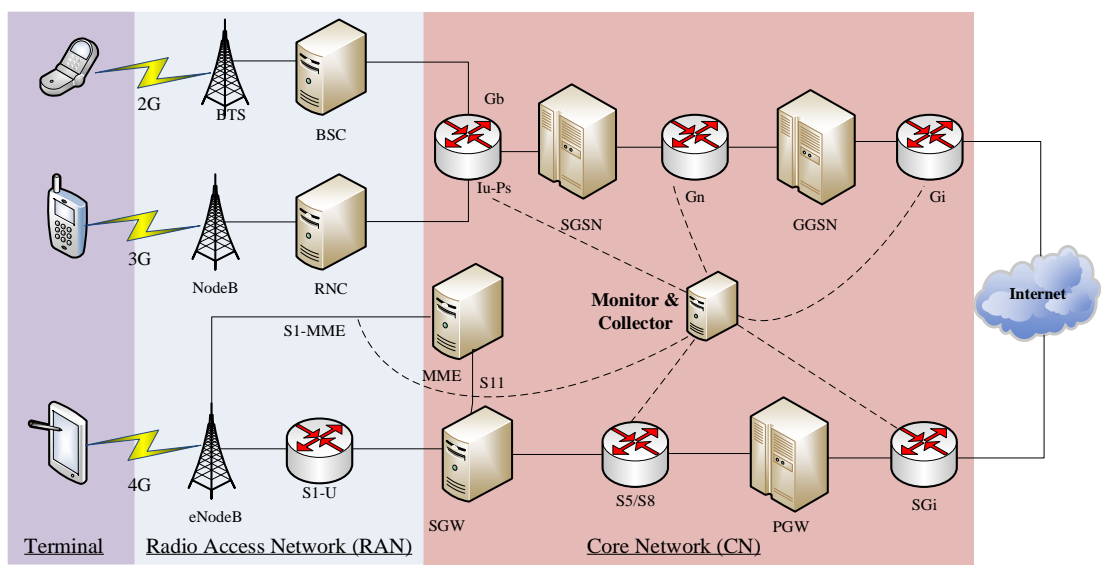

Figure 1.    Mobile Cellular Network Architecture and the data collecting points

Table 1    Data Categories in Mobile Cellular Networks

| Data Categories | specific item | Description |
|---|---|---|
| **Flow record data** | Data XDR | Flow-type data recording subscribers' data access attributes during a connection session |
| | Signaling XDR | Flow-type signaling information |
| **Network performance data** | Data for network KPI (Key Performance Indicator) | Evaluating the performance and resource utilization of air interface for both upload and download |
| | Measurement Report (MR) | Air interface channel quality and interference information |
| **Mobile terminal data** | Smartphone Data | Rich log file recording subscriber's services and terminal information |
| | IoT ( M2M) Data | Increasing rapidly with network scale |
| **Additional data** | User Profile | Billing information, subscribers data plan, etc |
| | Geographic Information | BS/Cell location, POI |

a)   **Flow Record Data**

A flow is a collection of packets between two end nodes defined by specific attributes, such as the well-known TCP/IP five-tuples. The flow record data in cellar networks are typically obtained through Deep Packet Inspection (DPI) and Deep Flow Inspection (DFI), which includes both data record and signaling record, in the form of XDR (call/transaction Detailed Record). The data records contain the main attributes during a data connection session, such as user id (e.g., IMSI, IMEI, MSISDN, etc), locations (e.g., Cell ID and its longitude and latitude), the timestamp of flow begin and end, the total number of packets and bytes, Uniform Resource Identifier (URI ) hostname and server IP, service types, etc. The flow record is perhaps the most important data describing the behavior of subscribers. For a medium-size city (1 million subscribers) in one week, the total flow record data is about 70TB (approximately $10^{15}$ records).

b)   **Network Performance Data**

Network performance data refers to aggregated data sets collected during a certain period (e.g., 5 minutes, 15 min or half an hour). Network performance data mainly include the Key Performance Indicator (KPI) data and measurement report (MR). KPIs are widely used in mobile cellular networks with the aim to evaluate the network performance and Quality of Service (QoS) delivered to users. KPIs pave the way for radio network quality, resource utilization and coverage of the wireless networks. Measurement reports are data that mainly contain information about channel quality. Measurement reports assist the network in making handover and power control decisions. Network performance data can grow to several Tbytes in one week for a medium-sized city.

c)   **Mobile Terminal Data**

With the development of the Internet of Things (IoT) and 4G cellular networks, a rapidly increasing number of connected devices are generating a large amount of data at terminals. Two typical categories are smartphone and IoT. Terminal data can be collected through a mobile APP [3].The data are from different layers, such as application layer, network layer and radio link layer. The data contains service characteristics, device information, and wireless network parameters such as international mobile subscriber identity, cell ID, signal strength, and download/upload rate, etc.

**d) Additional Data**

Some very important basic data information also exists in cellular networks, which can be summarized as subscriber profile and geographic information data. The subscriber profile includes a user's billing information and data plan. The geographic information data contains the location of the Cells/BSs, the point of interest (POI) information. Basic data information is relatively static compared with other data. However, these data contain information which is of vital importance for supporting data analysis.

## 2.2 Statistical Characteristics of Mobile Big Data

The datasets from real mobile networks have distinctive characteristics, featuring a wide variety of recording scale, temporal granularity, as well as several data types. For example, following the recording and acquisition of data from mobile networks, we can extract the location, mobility, proximity and the information of application usage. However, different from the traditional big data in computer networks, the wireless mobile big data has its typical statistical characteristics. In this paper, we study the following three main statistical characteristics for mobile data.

**a) Spatial-Temporal Distribution**

The data in mobile network can be partitioned to different granularities (for example, 5min, 15min, 1hour or even 1 day). Similarly, in the spatial domain, mobile data can be studied from a whole city, a typical area, one base station (BS) or even one cellphone. Heterogeneity and fluctuation commonly exist in both spatial and temporal distributions. Mobile traffic in different locations such as stadium, campus and dense residential areas exhibit different patterns in different time slots. Full utilization of such characteristics have already been enforced in the deployment of mobile networks.

**b) Data Aggregation Property**

Usually, the characteristics of aggregation property for mobile data are more important for network performance optimization. In the analysis of a metropolitan area in China, the total bytes distribution is highly uneven. For instance it is found that 20% subscribers will contribute more than 99% HTTP traffic [4]. From users' perspective, we focus more on the group user behavior rather than individual behavior [5]. For a given geographical area and a certain time period, group subscribers will probably request the same traffic, resulting in a similar traffic pattern. The aggregation features will be further used in predictive modeling to improve network performance.

**c) Social Correlations**

In wireless cellular networks, due to the social nature and habits of human-being, users close in the vicinity tend to exhibit similar habits, behavior and mobility rules [6]. For example, the social correlations of users lead to a larger scale of traffic correlation in both temporal and spatial domain, such as the autocorrelation and cross-correlation properties of mobile traffic. Thus, social correlations are a universal phenomenon in mobile data. How to make full use of social correlations is of great importance to benefit the mobile network.

## 3. Social Characteristics of Mobile Big Data

In current literature and applications, mobile big data has been studied and utilized from various perspectives. Instead of viewing mobile big data as a pure burden, in this section we investigate the potential performance gain of analyzing mobile big data from the perspective of social network analysis (SNA) [7]. SNA is an interdisciplinary academic field which emerged from sociology, statistics, and graph theory, which has been widely used to study relationship between individuals, groups, organizations, or even entire societies [8]. In this section, we explore the social characteristics of mobile big data with an emphasis on three aspects: users, base stations and applications.

### 3.1 Users Social Network

A user social network (USN) describes the social relationship between users and has received much attention recently in many fields. As for mobile data, several classic methods have been proposed to construct the user social network by using CDRs (Call Detailed Records), which contain calling/called information. In such cases, social graph construction approaches are based on using phone numbers (users) for the nodes, and call connection for the graph edges. In recent years, as more people use mobile devices to access the Internet—including online social networks such as Facebook and Twitter—we see a convergence of social and mobile networks. People tend to highly value the content recommended by friends or people with similar interests. Thus, the network edges can be easily defined in several ways, such as the number of contents shared between two subscribers, their common interests and the habits of each individual during specific time (weekday, weekend), etc. Once established, the USN can be used to detect community structures, understand true communication behavior, churn prediction and build a user-centric wireless network.

### 3.2 Base Stations Social Network

In mobile networks, it's difficult to obtain the accurate position of the users. However, the location of a base station or a cell tower can provide rough spatial information for wireless traffic which is sufficient from the perspective of a large metropolitan area. Here, we present the concept of a BS social network (BSSN) and construct one with collected traffic data from Hong Kong's LTE network.

As shown in Fig.2. Each node in a BSSN is a BS. Unlike a USN, the edges in a BSSN represent relationships rather than real social ties. To construct a BSSN, firstly the relationship between BSs is quantified with the BSs' traffic traces by using Pearson correlation coefficient. Pearson correlation coefficient estimates the strength of a linear relationship between two BSs, giving a value between +1 and -1 inclusive. Then, a planar maximally filtered graph is applied to filter the relationship to obtain a BSSN.

Several interesting work can be done based on a BSSN, such as link prediction, community detection, network evolution, and social influence analysis, but we mainly focus on a BSSN's community structure. "Community" refers to a subgraph structure within which nodes have a higher density of edges, whereas vertices between sub-graphs have a lower density. As shown in Fig.2, there exists four groups of BSs within which the connections are dense, while sparse between groups. Analytical results show that each community in a BSSN corresponds to a typical scenario with a common traffic pattern, which can be used to recognize typical traffic scenarios to adapt dynamic resource allocation.

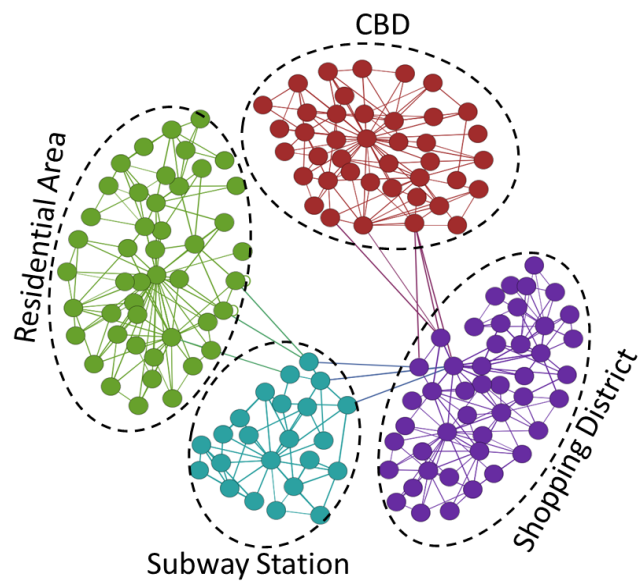

Figure 2. Base Station Social Network: BSs of the same color belong to the same community. Four communities including CBD, shopping district, residential area and subway stations are shown.

## 3.3 Apps Social Network

With the popularity of smartphones and mobile applications, many unique mobile applications continue to emerge, such as location-based services, mobile games, and mobile commerce applications. The data traces collected from mobile networks contains a large amount of applications in people's daily life. With the massive data we try to find the relations between the use of application. Using the similar method as constructing a BSSN, an App social network (ASN) is established with the real spatial-temporal cellular network traffic data collected from Shenzhen's 3G network, as shown in Fig. 3.

In an ASN, each node represents one typical App, and the edges indicate the strength of relationship among various Apps. The App with higher degree has a larger size and this App trends to be used with more other applications at a specific time. Apps of the same color belong to the same application community, such as video, Instant Messaging (IM, e.g., WeChat), office applications and others.

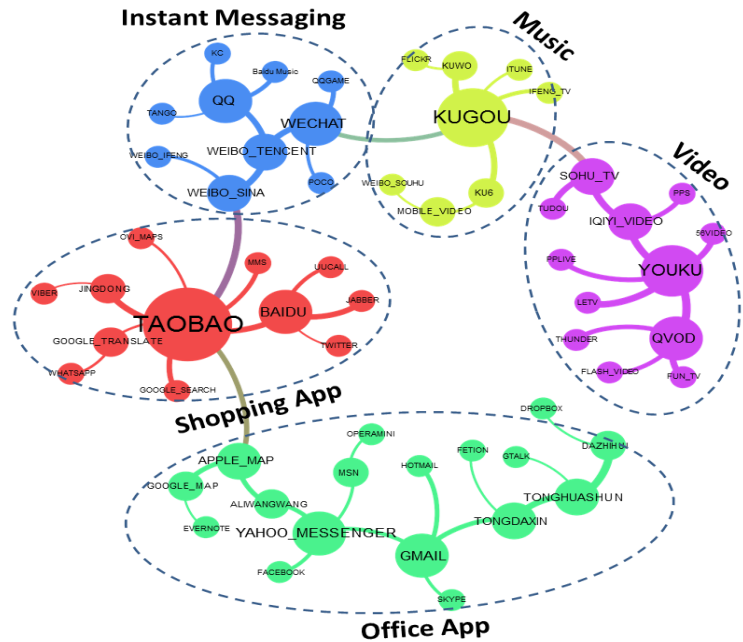

Figure 3.  App social networks. Each node represents one typical app, and the edges indicate the strength of relationships among various apps. The node (app) with higher degree has a larger size; apps of the same color belong to the same community, such as video, instant messaging, office applications, and others.

### 3.4 Interaction between Different Social Networks

With the vast mobile data collected from wireless cellular networks, USN, BSSN and ASN are established from the SNA perspective: USNs mainly reflect on the social ties between users and model their behavior characteristics, BSSNs study the traffic relationship of each BS, and ASNs represent the usage patterns of diverse apps. With the emergence of new mobile applications, we see the convergence of these three networks. For example, Pokemon Go (a game in which people use real-time maps to search for Pokemon characters) recently became very popular all over the world. Players in the same vicinity (within one BS) share the same live map and can cooperate with each other. Hence, this social network can be exploited to provide a quality experience by utilizing the app's data.

## 4. Conclusions

Our results span a wide range of subjects in the field of social computing for mobile big data; however, there are many open questions. First, new social relationships (besides USN, BSSN, and ASN) from mobile big data need to be explored. Second, the dynamics and evolution of mobile big data's social characteristics should be investigated. Third, how do we exploit the social characteristics from mobile big data to design and optimize practical cellular networks?

Along with the evolution of mobile network architecture and other technologies like virtual and augmented reality, social computing will be incorporated in all aspects of cellular networks. Operators and app developers face a great challenge in making full use of the available information.

## 5. Acknowledge

This work is supported by the National 973 Program under grant 2012CB316005, by the National Science Foundation of China (NSFC) under grant 61372114, 61571054 and 61631005, by the New Star in Science and Technology of Beijing Municipal Science & Technology Commission (Beijing Nova Program: Z151100000315077).